\documentclass[conference,letterpaper]{IEEEtran}

\usepackage{cite}
\usepackage{bm}
\usepackage{amsmath}
\usepackage{extarrows}
\usepackage{amssymb}
\usepackage{graphicx}
\usepackage{color}
\usepackage{enumerate}
\usepackage{bookmark}
\graphicspath{{figure}}

\usepackage[utf8]{inputenc}
\usepackage[T1]{fontenc}
\usepackage{url}
\usepackage{ifthen}
\usepackage{stfloats}

\newcommand{\mr}{\mathrm}

\newcommand{\BE}{\begin{equation}}
\newcommand{\EE}{\end{equation}}
\newcommand{\BS}{\begin{subequations}}
\newcommand{\ES}{\end{subequations}}
\renewcommand{\bf}{\bm}

\newtheorem{theorem}{Theorem}
\newtheorem{proposition}{Proposition}
\newtheorem{assumption}{Assumption}

\ifCLASSINFOpdf
\graphicspath{{figure/}}
\else
\fi
\begin{document}

\title{Capacity Optimality of AMP in Coded Systems}




\author{\IEEEauthorblockN{Lei~Liu\IEEEauthorrefmark{1}\IEEEauthorrefmark{2},  \emph{Member, IEEE},  Chulong~Liang\IEEEauthorrefmark{1}\IEEEauthorrefmark{3}, Junjie~Ma\IEEEauthorrefmark{4}, and~Li~Ping\IEEEauthorrefmark{1}, \emph{Fellow, IEEE}}\vspace{-3mm}\\\normalsize
           \IEEEauthorrefmark{1}Department of Electronic Engineering, City University of Hong Kong, Hong Kong\\
           \IEEEauthorrefmark{2}School of Information Science, Japan Institute of Science and Technology, Nomi 923-1292, Japan\\
           \IEEEauthorrefmark{3}State Key Laboratory of Mobile Network and Mobile Multimedia Technology, ZTE Corporation, Shenzhen, 518057, China\\
          \IEEEauthorrefmark{4}Institute of Computational Mathematics and Scientific/Engineering Computing, Chinese Academy of Sciences, China  
           }

\maketitle

\begin{abstract}
This paper studies a large random matrix system (LRMS) model involving an arbitrary signal distribution and forward error control (FEC) coding. We establish an area property based on the approximate message passing (AMP) algorithm. Under the assumption that the state evolution for AMP is correct for the coded system, the achievable rate of AMP is analyzed. We prove that AMP achieves the constrained capacity of the LRMS with an arbitrary signal distribution provided that a matching condition is satisfied. We provide related numerical results of binary signaling using irregular low-density parity-check (LDPC) codes. We show that the optimized codes demonstrate significantly better performance over un-matched ones under AMP. For quadrature phase shift keying (QPSK) modulation, bit error rate (BER) performance within 1 dB from the constrained capacity limit is observed. 
\end{abstract}

\textit{A full version of this paper is accessible at
\href{https://arxiv.org/pdf/1901.09559.pdf}{arXiv} (see \cite{Lei2019arXiv}).\footnotetext{The work was supported in part by the Japan Society for the Promotion of Science (JSPS) Kakenhi under Grant JP 21K14156, and in part by the University Grants Committee of the Hong Kong Special Administrative Region, China, under Grant CityU 11 216 518 and Grant CityU 11 209 519.}}



\section{Introduction}
Consider the problem of signal reconstruction for a large random matrix system (LRMS):
\BE\label{Eqn:linear_system}
\bf{y}=\bf{Ax}+\bf{n} 
\EE
where $\bf{A}$ is an $M\times N$ matrix with independent and identically distributed (IID) entries and $\bf{x}$ a length-$N$ vector with IID entries. Furthermore, we assume that the entries of $\bf{A}$ are Gaussian, but those of $\bf{x}$ are not necessarily Gaussian. 

In a special case when $\bf{x}$  is un-coded, if $\bf{x}$ is Gaussian, the optimal solution can be obtained using the standard linear minimum mean square error (MMSE) methods. Otherwise, the problem is in general NP hard \cite{Micciancio2001,verdu1984_1}. Approximate message passing (AMP), derived from belief-propagation (BP) with Gaussian approximation and first order Taylor approximation, has attracted extensive research interest for this problem \cite{Bayati2011, Donoho2009}. A basic assumption of AMP is that $\bf{A}$ is IID Gaussian (IIDG). This assumption will hold throughout this paper.

 AMP works by iterating between two local processors: namely, a linear detector (LD) and a non-linear detector (NLD). There is no matrix inversion involved, so its complexity is low. AMP has been studied for various signal processing and communication applications \cite{Ma2014denoising, Rush2017SSC, Liang2020, Barbier2017SSC, Liang2017CL}. Recently, it has been observed that AMP and its variations such as expectation propagation (EP) \cite{Cakmak2018, Minka2001} and orthogonal AMP (OAMP) \cite{Ma2016} can outperform the conventional Turbo linear MMSE (Turbo-LMMSE) in coded linear systems involving FEC coding \cite{Santos2017, MengVTC2015, MaTWC}. Most works on AMP in coded systems are simulation based. There is still a lack of rigorous analysis on the information theoretical limits of AMP in coded systems. 

In this paper, we discuss the LRMS in \eqref{Eqn:linear_system} with FEC coding. The receiver is a variation of AMP with NLD formed by an \emph{a posteriori probability} (APP) decoder. For convenience of discussions, we define two classes of optimality for a receiver.
\begin{itemize}
  \item A receiver is MMSE-optimal if it can achieve MMSE when $\bf{x}$ is an IID sequence.
  \item A receiver is information theoretically optimal if it can achieve error free performance when $\bf{x}$ is coded with a rate which equals to the mutual information $I(\bf{x}; \bf{y})$. 
\end{itemize}

The state evolution (SE) technique of AMP was originally derived to track the mean square error (MSE) in AMP during iterative processing. SE involves a scalar recursion of the transfer functions of LD and NLD. It has been shown via SE analysis that AMP can achieve MMSE asymptotically in the un-coded case when the transfer functions of LD and NLD have only one fixed-point \cite{ Tulino2013, Barbier2017arxiv, Reeves_TIT2019}. In this paper, we will show via SE analysis that AMP is information theoretically optimal, while the conventional methods, such as the well-known Turbo-MMSE algorithm  \cite{Lei20161b,Yuan2014}, are not.  

Our discussions are based on the following background works: (i) the I-MMSE relationship between mutual information and MMSE  \cite{Guo2005}, (ii) the area property of iterative decoding systems \cite{Bhattad2007}, and (iii) the MMSE-optimality of AMP \cite{Tulino2013, Barbier2017arxiv, Reeves_TIT2019}. Similarly to \cite{Yuan2014, Lei20161b }, the performance of AMP can be optimized by matching the transfer functions of LD and decoder. The achievable rate can be analyzed using an area property similar to that for low density parity check (LDPC) decoders \cite{Yuan2014, Lei20161b}. We find that perfect matching is impossible for AMP: there is an inherent gap between the two transfer functions. Interestingly, AMP is still information theoretically optimal despite this gap, in the sense that its achievable rate can approach the mutual information $I(\bf{x}; \bf{y})$. The following are the main contributions of this paper.
\begin{itemize}
  \item We show that the constrained capacity of a coded LRMS with an arbitrary input distribution (Gaussian or non-Gaussian) can be graphically interpreted as the area determined by the transfer functions of LD and MMSE NLD of an AMP. We establish an area property for AMP and derive its achievable rate under a matching condition. We prove that this achievable rate equals to the constrained capacity of an LRMS derived in \cite{Barbier2017arxiv,  Reeves_TIT2019}, thereby showing the potential information theoretic optimality of AMP in coded linear systems.  
  
  \item  We develop a matching strategy for AMP. We provide numerical results to demonstrate the efficiency of the matching strategy for binary signaling. These findings provide a promising direction to significantly enhance the performance of coded linear systems.  
   
\end{itemize}

\section{Preliminaries} 
{In this section, we briefly outline the area property, the capacity of an LRMS, and the AMP algorithm.

\subsection{Area Property for SISO-AWGN Channel}\label{Sec:SISO_area}
A SISO-AWGN channel is defined as \BE\label{Eqn:SISO_AWGN}
y=\sqrt{\rho}x+z,
\EE
where $z\sim \mathcal{CN}(0,1)$, $\rho$ denotes the signal-to-noise-ratio (SNR), $x\sim P_X(x)$ and $P_X(x)$ is an arbitrary distribution on a constellation $\mathcal{S}$. The MMSE of \eqref{Eqn:SISO_AWGN} is denoted as
\BE
\omega_{\cal S}(\rho) \equiv \mr{mmse}(x|\sqrt{\rho}x+z, x\sim P_X(x)).
\EE

The following theorem, proved in \cite{Guo2005}, establishes the connection between MMSE and the capacity given $P_X(x)$ for a SISO-AWGN channel.

\begin{theorem}[Scalar I-MMSE]\label{Lem:S-I-MMSE}
Let SNR$=\rho^*$. The capacity of a SISO-AWGN channel equals to the area under  $\omega_{\cal S}(\rho)$ from $\rho=0$ to $\rho=\rho^*$, i.e., 
\BE\label{Eqn:C_mmse}
\!\!C_{\rm SISO}(\rho^*) \!=\!\! I({x}; \sqrt{\!\rho^*}x\!+\!z) = \!\!\int_{0}^{\rho^*} \omega_{\cal S}(\rho) d\rho.
\EE
\end{theorem}
 
The following is an instance of Theorem \ref{Lem:S-I-MMSE}. 
 
 \emph{Code-Rate-MMSE Lemma \cite{Bhattad2007}:} Let the code length be $N$ and code rate $R_{\cal C} =K/N$. We treat the code-book $\bf{{\mathcal{C}}}=\{\bf{c}_1,\cdots,\bf{c}_{2^K}\}$ as a uniformly distributed $N$-dimension constellation with ${2^K}$ discrete points. When SNR$\to \infty$, the capacity per length-$N$ code block approaches to the entropy of $\bf{{\mathcal{C}}}$, i.e., $\log(2^K)=K$. The entropy per dimension is $K/N$. Hence,  
\BE\label{Eqn:R_mmse}
R_{\cal C} =  \int_{0}^{\infty}\omega_{\mathcal{C}}(\rho)d\rho =K/N,
\EE
where  $ \omega_\mathcal{C}(\rho) \!\equiv\! \tfrac{1}{N}\mr{mmse}(\bf{x}|\sqrt{\rho}\bf{x}\!+\!\bf{z},{\bf{x}\!\in \!{\bf{{\mathcal{C}}}}})$.

\subsection{LRMS Capacity}
Return to the LRMS in \eqref{Eqn:linear_system}: $\bf{y}=\bf{Ax}+\bf{n}$,
where $\bm{y}\!\in\!\mathbb{C}^{M\!\times\!1}$ is a vector of observations, $\bf{A}\!\in\!\mathbb{C}^{M\!\times\! N}$ an IIDG matrix with $A_{ij}\sim \mathcal{CN}({0},1/M)$, $\{x_i\sim P_X(x), \forall i\}$, and $\bm{n}\!\sim\!\mathcal{CN}(\mathbf{0},\sigma^2\bm{I}_M)$ a vector of Gaussian additive noise samples. Fig. \ref{Fig:model}(a) shows 
a modulated LRMS. In this paper, we consider a large system with $M,N\to\infty$ and a fixed $\beta=N/M$. The transmit SNR is defined as $snr =  \sigma^{-2}$.
We assume that $\bf{A}$ is known at the receiver, but unknown at the transmitter. 

 \begin{figure}[t]
  \centering
  \includegraphics[width=6cm]{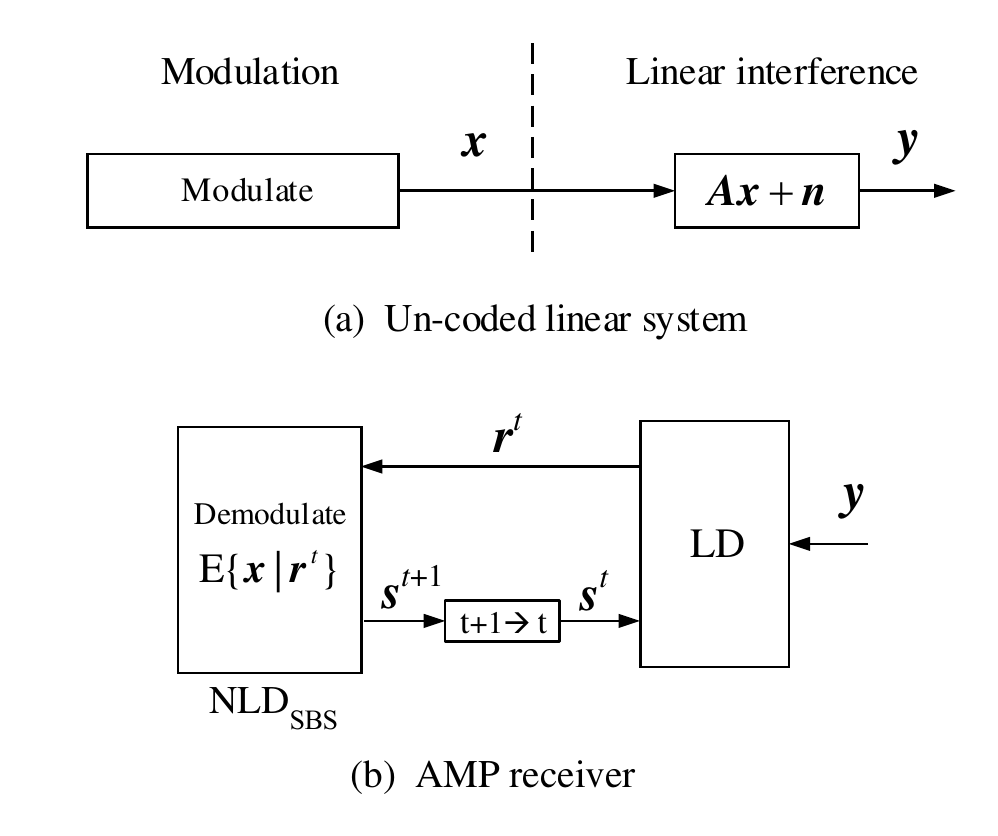}\\ \vspace{-0.2cm}
  \caption{Un-coded LRMS: transmitter and AMP receiver, where ``Demodulate'' and LD in (b) correspond to ``Modulate'' and ``$\bf{A}\bf{x}+\bf{n}$'' in (a) respectively.}\label{Fig:model}\vspace{-0.2cm}
\end{figure}

The constrained capacity of an LRMS given $P_X(x)$ was proved in \cite{Barbier2017arxiv, Reeves_TIT2019}. 
  

\begin{theorem}[Capacity]\label{Pro:dis_cap}   
Assume that the signal distribution $P_X(x)$ satisfies the single-crossing property, i.e., $\zeta \!=\!\beta \cdot snr \cdot \omega\big(1/[\beta(1+\zeta)]\big)$ has exactly one positive fixed point $\zeta^*$. Then, the capacity of the LRMS in \eqref{Eqn:linear_system} is given by \vspace{-0.1cm}
\BE\label{Eqn:dis_cap}
C  \!=\!  \beta^{-\!1}\!\big[\! \log({1\!+\!\zeta^*})\!-\!{\zeta^*}\!/\!({1\!+\!\zeta^*}) \!\big] \!+ C_{\rm SISO} \big(snr/(1\!+\!\zeta^*)\!\big),  
\EE 
where $C_{\rm SISO}(\cdot)$ is defined in \eqref{Eqn:C_mmse}.
\end{theorem}

In \cite{Lei2019arXiv}, we provide a concise derivation of the  LRMS capacity, using the properties of AMP. 


\subsection{Overview of AMP}\label{Sec:AMP_uncoded}
AMP \cite{Donoho2009}  finds an approximate MMSE solution to the problem in \eqref{Eqn:linear_system} using the following iterative process (initialized with $t=0$ and $\bf{s}^0=\bf{r}^0_{\mr{Onsager}}=\bf{0}$): 
\BS\label{Eqn:AMP}\begin{align}
 \mathrm{LD:}\;\; & \bf{r}^t\!=\! f(\bf{s}^t) \!\equiv \! \bf{s}^t \!+\! \bf{A}^H(\bf{y}\!-\!\bf{A}\bf{s}^t) \!+ \!\bf{r}^t_{\mr{Onsager}},\label{Eqn:LD}\\
\mathrm{NLD_{SBS}}: \;\; &\bf{s}^{t+1} = \eta(\bf{r}^{t})\equiv \mr{E}\{\bf{x}|\bf{r}^{t}\},\label{Eqn:NLD}
\end{align}\ES
where $\eta(\bf{r}^{t})$ is a symbol-by-symbol (SBS) MMSE demodulate function, and $\bf{r}^t_{\mr{Onsager}}$ is an ``Onsager term''  defined by $\bf{r}^t_{\mr{Onsager}}\!=\!\beta \langle\eta'(\bf{r}^{t-1})\rangle (\bf{r}^{t-1}\!-\!\bf{s}^{t-1})$ \cite{Donoho2009}. Fig. \ref{Fig:model}(b) is a graphical illustration of AMP, where the linear detector (LD) and non-linear detector (NLD) correspond to \eqref{Eqn:LD} and \eqref{Eqn:NLD} respectively. 
We define the errors as
\BE\label{Eqn:errors}
\bf{h}^t \equiv \bf{r}^t -\bf{x} \quad {\rm and} \quad  \bf{q}^t \equiv \bf{s}^t -\bf{x}.
\EE 
Let $\rho^t$ be the signal-to-interference-plus-noise-ratio (SINR) for $\bf{r}^t$ and $v^t$ the MSE for $\bf{s}^t$:
\BE\label{Eqn:rho_v}
  \rho^t \equiv N\big[{\rm E}\big\{\|\bf{h}^t\|^2\}\big]^{-1},\qquad
  v^t \equiv  \tfrac{1}{N}{\rm E}\big\{\|\bf{q}^t\|^2\big\}.
\EE
The following theorem summarizes the findings in \cite{Bayati2011}.  

\begin{theorem}\label{Pro:SE}
Let $M,N\to\infty$ with a fixed $\beta=N/M$. For AMP, $\bf{h}^t$ defined in \eqref{Eqn:errors} can be modeled by a sequence of IIDG samples independent of $\bf{x}$. The LD and NLD of AMP can be characterized by the following transfer functions \cite{Bayati2011}
 \BS \begin{align}
\mathrm{LD:}& \quad \rho^t  =\phi(v^t) = ({\beta v^t + \sigma^2})^{-1},\label{Eqn:LD_form}\\
\mathrm{NLD_{SBS}:}& \quad  v^{t+1}= \omega_{\cal S}(\rho^t), \label{Eqn:NLD_MMSE}
\end{align}\ES
where $\omega_{\cal S}(\cdot)$ is the MMSE function given in \ref{Sec:SISO_area}. 
\end{theorem}
  
\begin{figure}[t] 
  \centering
  \includegraphics[width=6cm]{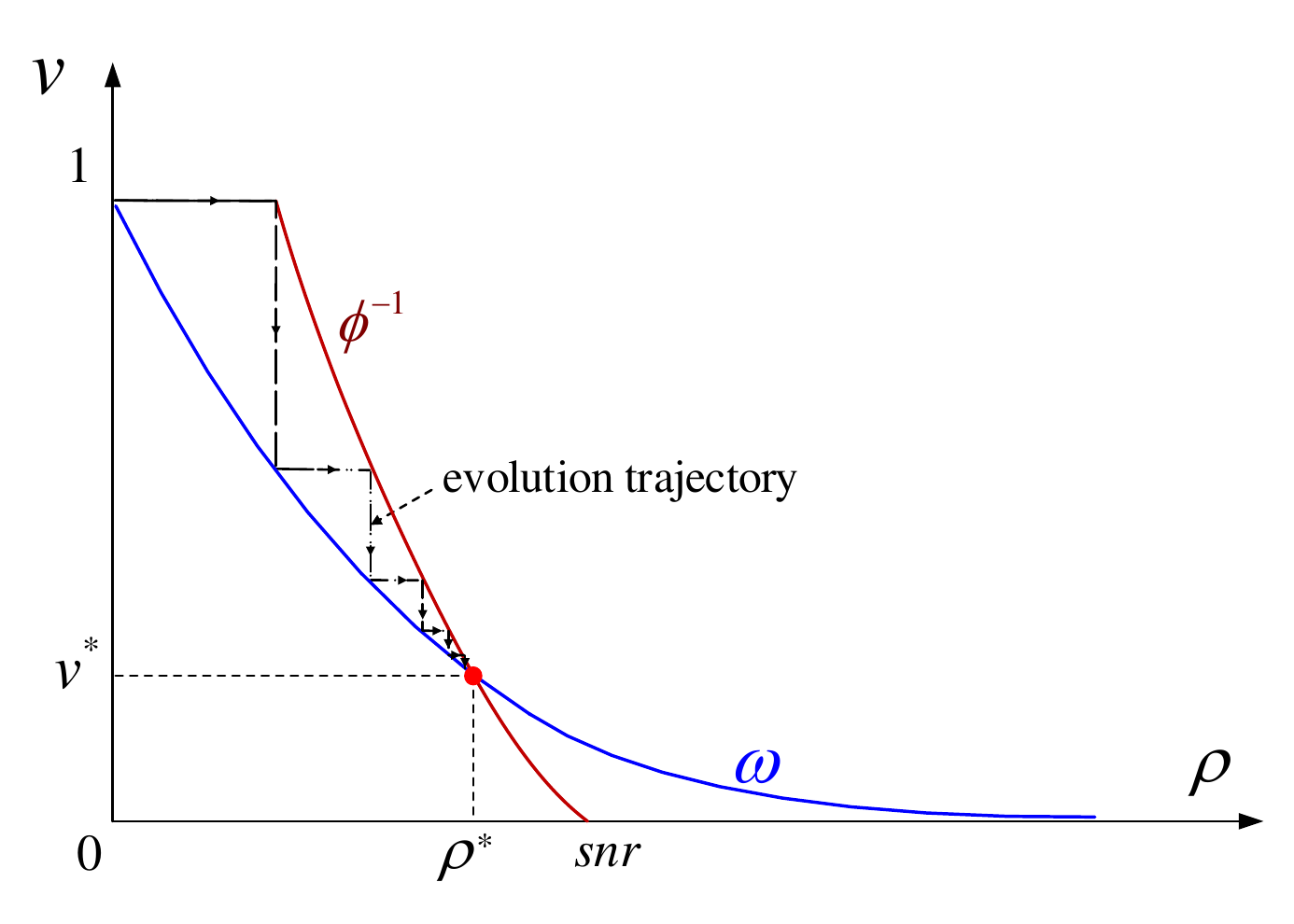}\\
  \caption{Graphical illustration of an  AMP, where $\phi^{-1}$ is the inverse of $\phi$ given in \eqref{Eqn:LD_form} and $\omega$ is defined in \eqref{Eqn:NLD_MMSE}. The iterative process of AMP is illustrated by the evolution trajectory, and the fixed point $(\rho^*, v^*)$ gives the MMSE. From \eqref{Eqn:LD_form}, we have $\phi(0)=snr$.}\label{Fig:TF_chart}
\end{figure}

{\begin{assumption}\label{Pro:SCP}
There is exactly one fixed point for $\omega_{\cal S}(\rho) = \phi^{-1}(\rho)$ for $\rho>0$, where $\phi^{-1}(\cdot)$ is the inverse of $\phi(\cdot)$.
\end{assumption}}

It is proved in \cite{Lei2019arXiv} that Assumption \ref{Pro:SCP} rigorously holds for Gaussian signaling.  Fig. \ref{Fig:TF_chart} provides a graphical illustration of Assumption \ref{Pro:SCP}. The evolution trajectory of AMP converges to a unique fixed point $(\rho^*, v^*)$. The following theorem was first established in \cite{Tulino2013} via replica method, and then was rigorously proved in \cite{Barbier2017arxiv, Reeves_TIT2019}.

\begin{theorem}[MMSE Optimality]\label{Lem:mmse}
Let $\hat{{\bf{x}}}_{\mr{MMSE}}=\mr{E}\{\bf{x}|\bf{y},x_i\!\sim\! P_X(x), \forall i\}$ be the conditional mean of $\bf{x}$ given $\bf{y}$ and $\{x_i\!\sim\! P_X(x), \forall i\}$ and suppose  that {Assumption \ref{Pro:SCP} holds.} Then 
\BE\label{Eqn:snr_rho}
v^*=\tfrac{1}{N}\mr{E}\big\{ \|\bf{x}\!-\!\hat{\bf{x}}_{\mr{MMSE}}\|^2 \big\},
\EE
 i.e., AMP converges to the MMSE of the un-coded LRMS.
\end{theorem}

\section{Capacity Optimality of AMP}\label{Sec:Cap_Opt} 
In this section, we investigate the achievable rate of the iterative AMP receiver with FEC decoding.

\subsection{Coded System Model and AMP}
\begin{figure}[b!] 
  \centering
  \includegraphics[width=6.5cm]{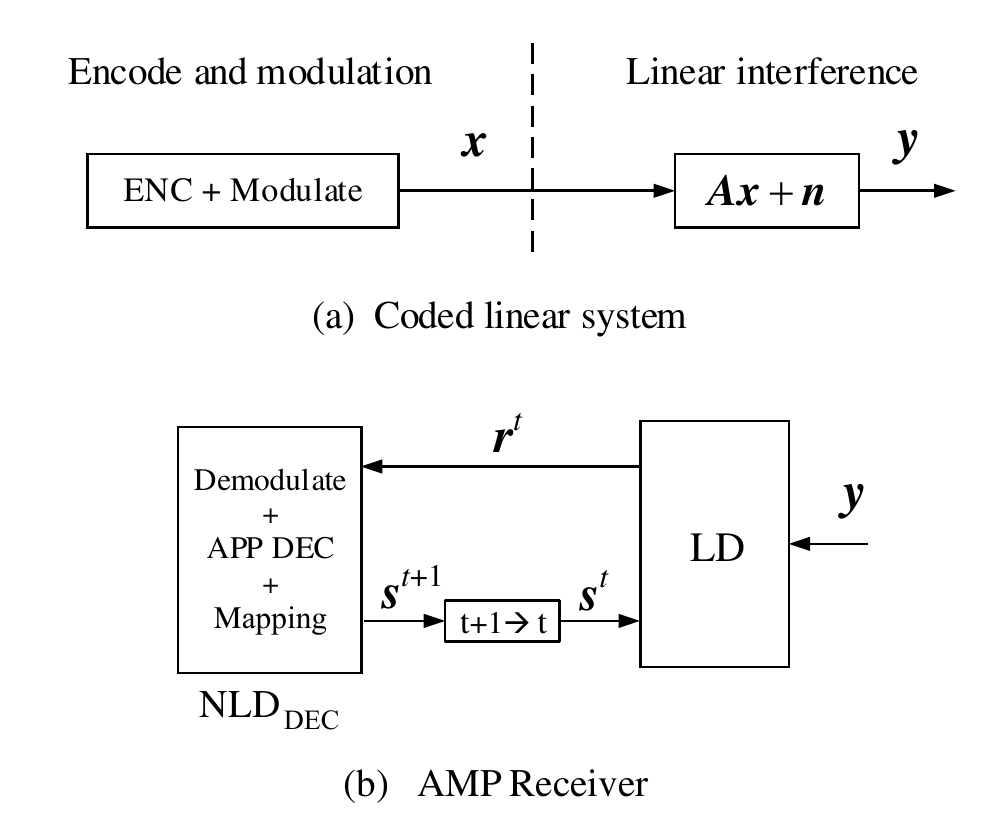}\\\vspace{-2mm}
  \caption{Coded linear system: Transmitter and AMP.   ``APP DEC'' (\emph{a-posteriori} probability decoding), ``Demodulate'' and LD in (b) correspond to ``ENC'' (encode),  ``Modulate'' and ``$\bf{A}\bf{x}+\bf{n}$'' in (a) respectively.}\label{Fig:model_coded} 
\end{figure} 
Fig. \ref{Fig:model_coded}(a) shows an LRMS involving FEC coding. We write $\bf{x}\in \bf{{\mathcal{C}}}$ for coded $\bf{x}$. The other conditions are the same as that in Fig. \ref{Fig:model}. We focus on the AMP receiver in Fig. \ref{Fig:model_coded}(b).

\underline{\emph{AMP:}} Initialized with $t\!=\!0$ and $\bf{s}^0\!=\!\bf{r}^0_{\mr{Onsager}}\!=\!\bf{0}$,
\BS\label{Eqn:Turbo-AMP}\begin{align}
&\mathrm{LD:}\quad\; \bf{r}^t\!=\! f(\bf{s}^t) \!\equiv \! \bf{s}^t + \bf{A}^H(\bf{y}\!-\!\bf{A}\bf{s}^t) + \bf{r}^t_{\mr{Onsager}}, \\
&\mathrm{NLD_{DEC}:} \;\; \bf{s}^{t+1} = \eta_{\cal C}(\bf{r}^{t})\equiv \mr{E}\{\bf{x}|\bf{r}^{t}, {\bf{x}\in {\bf{{\mathcal{C}}}}}\}.\label{Eqn:Turbo-NLD}
\end{align}\ES
The symbol-wise NLD in \eqref{Eqn:AMP} of AMP is replaced by an \emph{a-posteriori} probability (APP) decoder in \eqref{Eqn:Turbo-AMP} for coded $\bf{x}$.

Theorem \ref{Pro:SE} gives the IIDG property for AMP for un-coded $\bf{x}$. The discussions in this paper are based on the following assumption for coded $\bf{x}$. 

\begin{assumption}\label{Pro:SE_new}
Theorem \ref{Pro:SE} still holds for AMP for coded $\bf{x}$, i.e., $\bf{h}^t$ is IIDG and independent of $\bf{x}$ and LD and NLD of AMP can be characterized by   
\BS\label{Eqn:Turbo_AMP_SE}\begin{alignat}{2}
  \mathrm{LD:}  &\quad &&\rho  =\phi(v), \\
  \mathrm{NLD_{DEC}:}  &\quad  &&v\!= \!\omega_\mathcal{C}(\rho)\!\equiv\! \tfrac{1}{N}\mr{mmse}(\bf{x}|\sqrt{\rho}\bf{x}\!+\!\bf{z},{\bf{x}\!\in\! {\bf{{\mathcal{C}}}}}).\label{Eqn:NLD_MMSE_coded}  
 \end{alignat}
\ES
\end{assumption} 

The $\phi(v)$ in AMP is the same as that in AMP, and $\omega_\mathcal{C}(\rho)$ depends on the code constraint.

\subsection{Area Property and Capacity Optimality of AMP}\label{Sec:area_AMP} 
In the un-coded case in \eqref{Eqn:AMP}, AMP converges to a fixed $(\rho^*,v^*)$ in Fig. \ref{Fig:TF_chart}. Detection is not error free as $v^*>0$. In the coded case, it is possible to achieve error-free detection using a properly designed $\omega_\mathcal{C}(\rho)$. As illustrated in Fig. \ref{Fig:track}, the key is to create a detection tunnel that converges to $v=0$, implying zero error rate.  There should be no fixed point between $\omega_\mathcal{C}(\rho)$ and $\phi^{-1}({\rho })$, since otherwise the tunnel will be closed at $v>0$. This requires that 
 \BS\BE\label{Eqn:error-free}
\omega_{{\mathcal{C}}}(\rho) <\phi^{-1}(\rho), \;\;\;  {\rm for} \;\;0\le\rho\le snr.
 \EE
Also, by definition, the MMSE $\rm NLD_{DEC}$ in \eqref{Eqn:NLD_MMSE_coded} should achieve an MSE lower than that of a SBS detector, i.e.,
\BE\label{Eqn:coding_gain}
\omega_{{\mathcal{C}}}(\rho) < \omega_{\cal S}(\rho),  \;\;\;  {\rm for} \;\; \rho \geq 0.\vspace{-0.1cm}
\EE\ES
Combining \eqref{Eqn:error-free} and \eqref{Eqn:coding_gain}, we obtain a necessary and sufficient condition for AMP to achieve error-free detection:
\BS\label{Eqn:upper_bound}\BE
\omega_{\mathcal{C}}({\rho})< \omega_{\mathcal{C}}^*({\rho}),\;\;\; {\rm for} \;\; 0 \leq {\rho } \leq snr,
\EE
where
\BE \label{Eqn:w_star}
 \omega_{\mathcal{C}}^*({\rho})= \min \{\omega_{\cal S}(\rho),  \phi^{-1}({\rho }) \}.
\EE\ES

\begin{figure}[t]
  \centering
  \includegraphics[width=6.5cm]{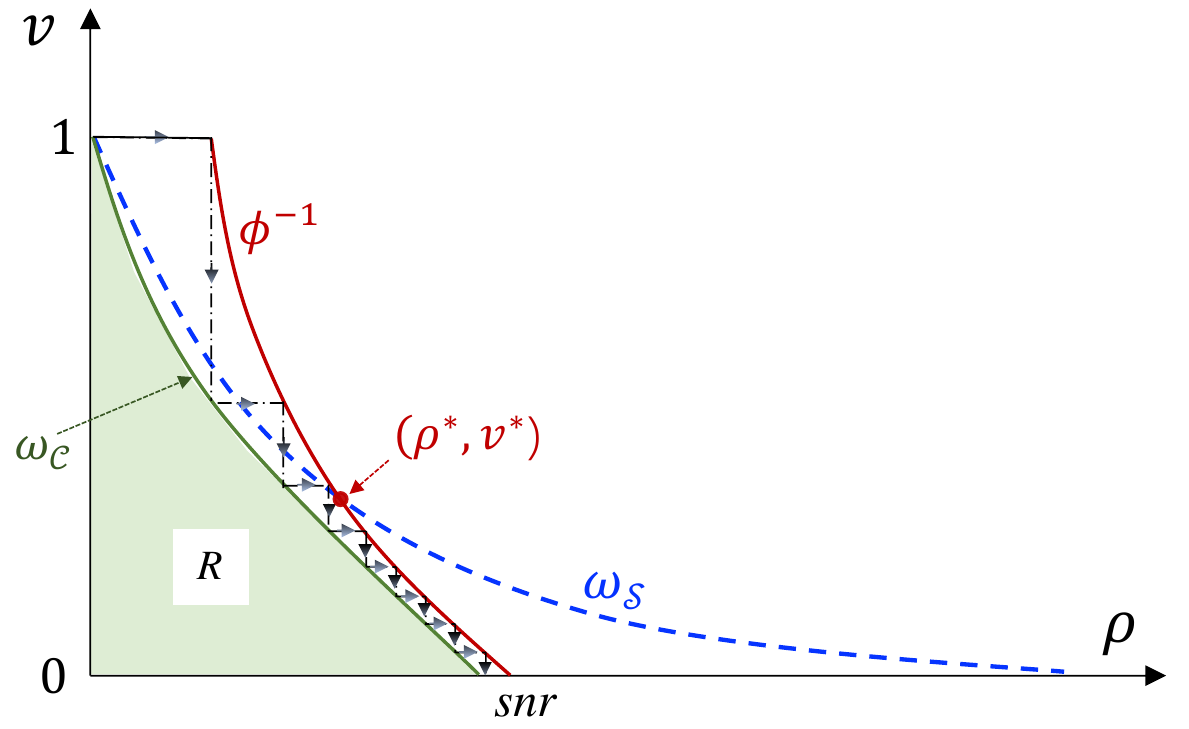}\\
  \caption{Graphical illustration of AMP, where $\omega_{\mathcal{S}}$ is a demodulation function (un-coded case) and $\omega_{{\mathcal{C}}}$ is a transfer function of a decoder (coded case). The iterative process of AMP is illustrated by the evolution trajectory between $\phi^{-1}$ and $\omega_{{\mathcal{C}}}$. }\label{Fig:track}
\end{figure} 

\begin{proposition}\label{The:area_LRMS}
Suppose that Assumption \ref{Pro:SCP} holds.  Then the constrained capacity of an LRMS with the given $\mathcal{\bf{S}}$ is  
\BS\label{Eqn:dis_cap_new2}\BE
C=A_{\omega_{\mathcal{C}}^*},
\EE 
where $A_{\omega_{\mathcal{C}}^*}$  is the area covered by $\omega_{\mathcal{C}}^*$, i.e.,  
\begin{align}
A_{\omega_{\mathcal{C}}^*}  =    \beta^{-1}[\rho^{*}/snr\!-\!\log(\rho^{*}/snr)\!-\!1] \!+\!\! \int_0^{\rho^{*}} \!\!\! \omega_{\cal S}(\rho) d \rho.
\end{align} \ES
\end{proposition}
\begin{IEEEproof} 
See APPENDIX \ref{APP:Consistency}.
\end{IEEEproof}

Combining \eqref{Eqn:R_mmse}, \eqref{Eqn:upper_bound} and \eqref{Eqn:dis_cap_new2}, we obtain the capacity optimality of AMP below.
\begin{theorem}[Capacity Optimality]\label{The:cap_opt}
Assume that Assumptions \ref{Pro:SCP} and \ref{Pro:SE_new} hold and AMP converges to $v=0$. Then, 
\BE
R_{\cal C}\to C,
\EE 
if $\omega_{\mathcal{C}}(\rho)\to \omega_{\mathcal{C}}^*(\rho)$ in $[0, snr]$.
\end{theorem}
 
Fig. \ref{Fig:Area_A} gives a graphical illustration of Theorem \ref{The:cap_opt}.  Note that Theorem \ref{The:cap_opt} is based on a matching condition:
\BE\label{Eqn:C2Copt}
\omega_{\mathcal{C}}(\rho)\to \omega_{\mathcal{C}}^*(\rho).
\EE
A proof for the existence of a code achieving \eqref{Eqn:C2Copt} can be found in \cite{Lei2019arXiv} for Gaussian signaling. For other signaling, the existence of such a code is a conjecture only.  
 
\begin{figure}[t] 
  \centering
  \includegraphics[width=6.5cm]{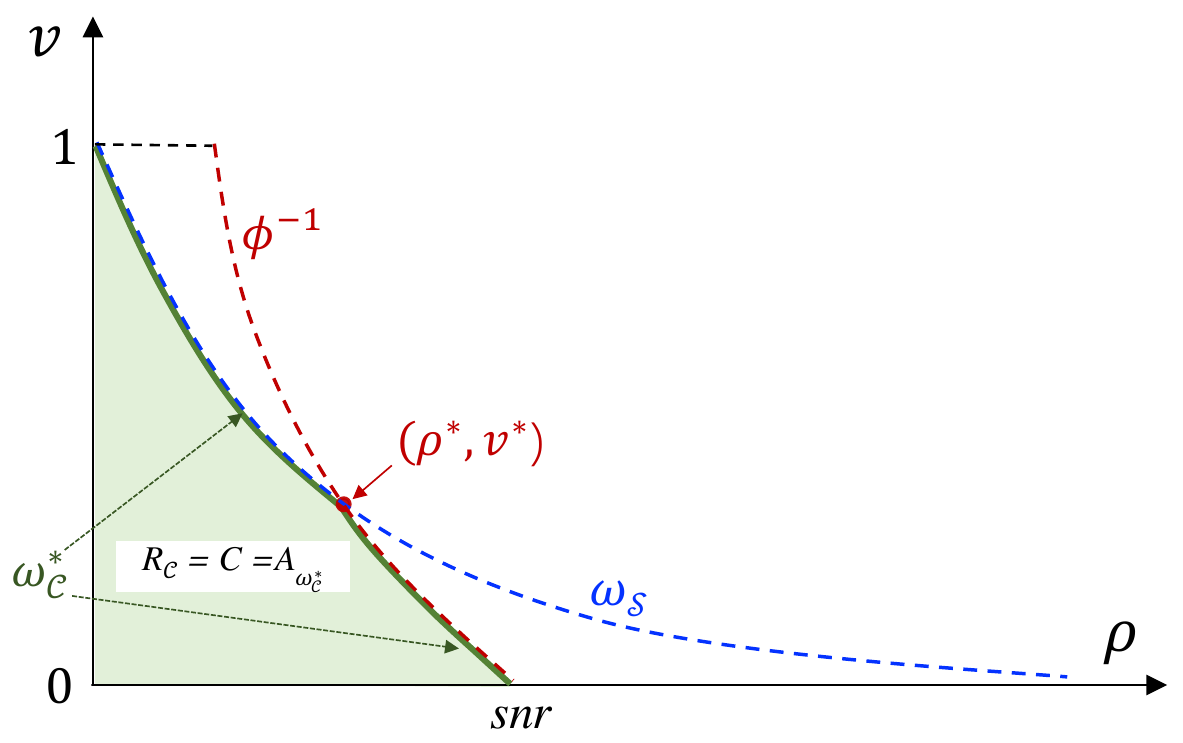}\\
  \caption{Graphical illustration of the capacity, the maximum achievable rate of AMP and the optimal transfer function of decoder. The maximum achievable rate of AMP equals to the capacity, which is the area covered by  $\omega_{\mathcal{C}}^*$. }\label{Fig:Area_A}
\end{figure}

\subsection{Rate Comparison with Turbo-LMMSE} 
It is proved in \cite{Lei20161b,Yuan2014} that Turbo-LMMSE is capacity achieving for Gaussian signaling. In the following, we show that Turbo-LMMSE is sub-optimal for non-Gaussian signaling.

{The main difference between AMP and Turbo-LMMSE is as follows. To avoid the correlation problem in the iterative process, Turbo-LMMSE uses extrinsic local processors (e.g. an extrinsic LD and an extrinsic decoder), while AMP uses an ``Onsager"-term.}
 

Assume that the transfer functions of the detector and the decoder in Turbo-LMMSE are matched. The achievable rate of Turbo-LMMSE is given in \cite{Yuan2014} 
\BE
R_{\mr{LMMSE}}=\log|\mathcal{S}|-\int_0^{+\infty} \omega_\mathcal{S}(\rho+\phi(\omega_\mathcal{S}(\rho))) d\rho.
\EE


Fig. \ref{Fig:Rate_AMP_Turbo} shows the capacity and the achievable rates of AMP and Turbo-LMMSE. The capacity for Gaussian signaling is achieved by both AMP and Turbo-LMMSE. For QPSK, 8PSK and 16QAM modulation, the achievable rate of AMP equals to the capacity when {Assumption \ref{Pro:SCP} holds}, while Turbo-LMMSE always has rate loss. Similar results can be obtained for other non-Gaussian signaling. In addition, the gap between AMP and Turbo-LMMSE increases with $\beta$. This gap $\to0$ when $\beta\to0$. {The reason why Turbo-LMMSE has performance loss is that extrinsic update leads to performance loss for non-Gaussian signal processing, which was first pointed out in \cite{MaTWC}. 

\begin{figure}[t]
  \centering
  \includegraphics[width=8.cm]{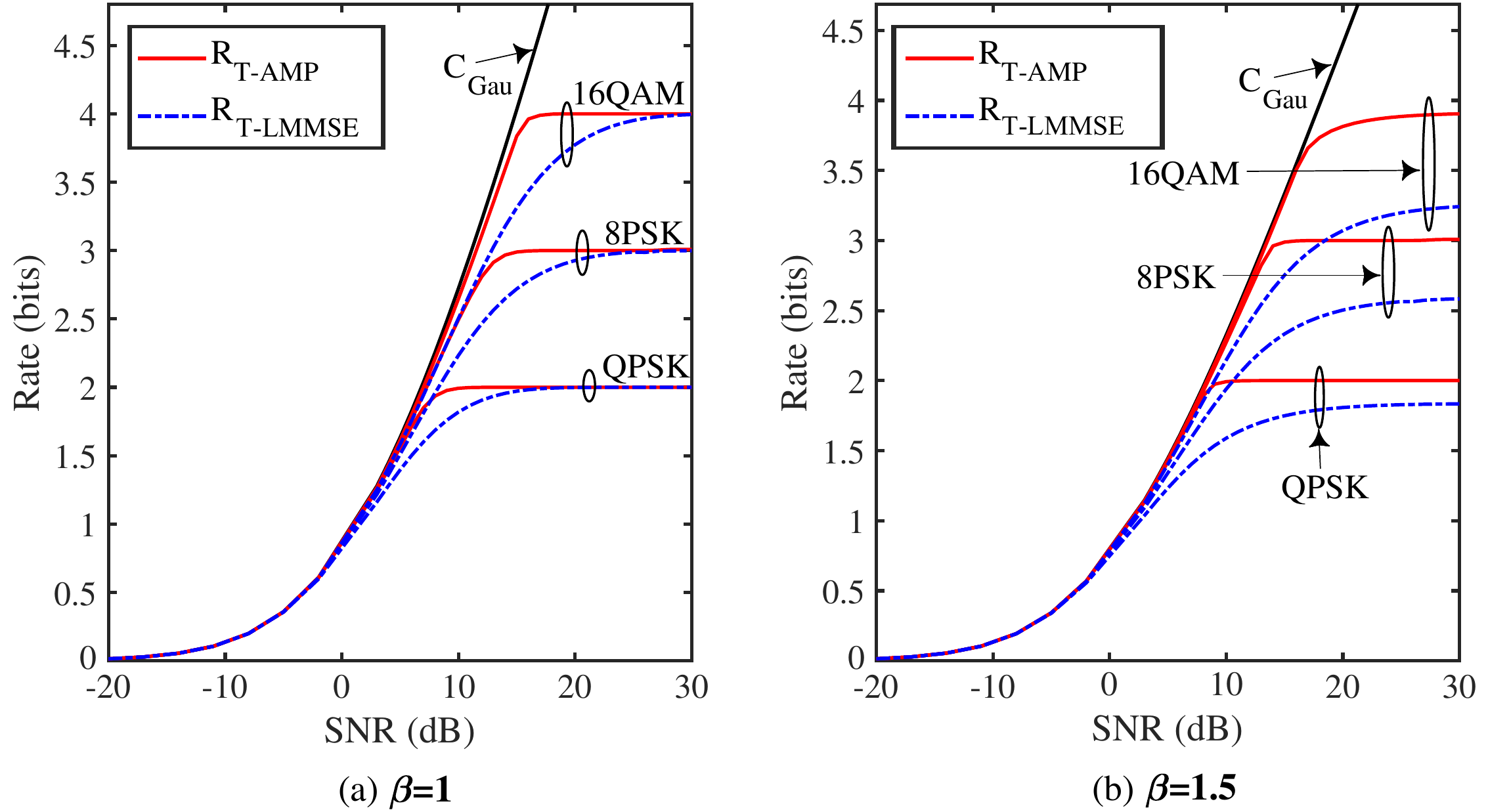}\\
  \caption{Comparison between the capacity and the achievable rates of AMP and Turbo-LMMSE of an LRMS with $\beta=N/M=\{1, 1.5\}$, where  $C_{\rm Gau}$ denotes the Gaussian capacity and also the achievable rates of AMP and Turbo-LMMSE with Gaussian signaling, $R_{\mr{T-AMP}}$ and $R_{\mr{T-LMMSE}}$ respectively denote the achievable rates of AMP and Turbo-LMMSE with QPSK, 16QAM and 8PSK modulations.}\label{Fig:Rate_AMP_Turbo}
\end{figure}

\subsection{Rate Comparison with Cascading AMP and Decoding}\label{Sec:AMP_DEC}  
We define a cascading AMP and decoding (AMP-DEC) scheme \cite{Guo2005random, Tanaka2002} as follows. We run AMP until it converges. The result is used by decoder. There is no iteration between AMP and the decoder. The achievable rate of AMP-DEC is  
\BE\label{Eqn:dis_ach}
R_{\mr{AMP-DEC}}=   \int_0^{{{\rho}}^{*}} \!\!\! \omega_\mathcal{S}(\rho) d \rho.
\EE

 \begin{figure}[t]
  \centering
  \includegraphics[width=8.cm]{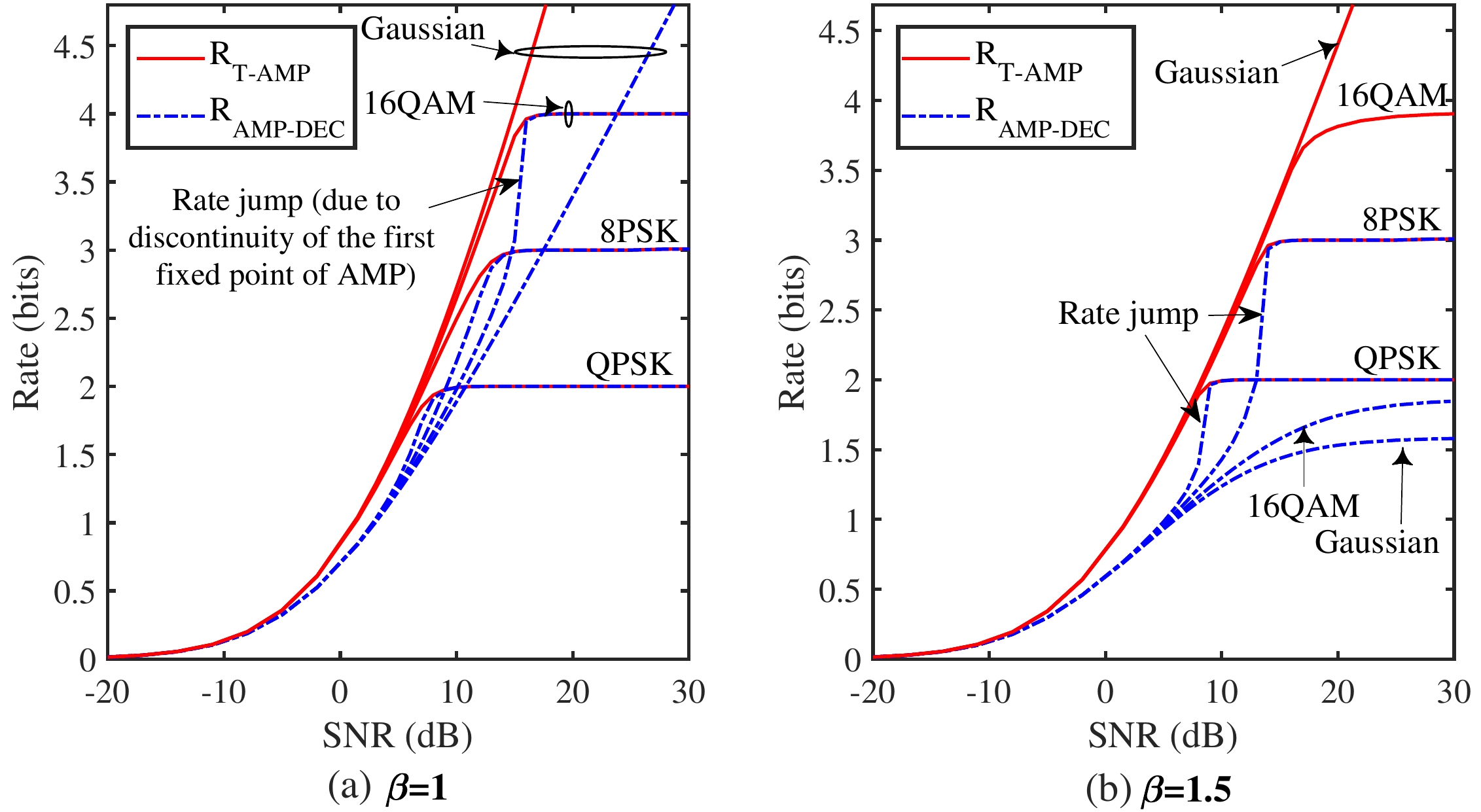}\\
  \caption{Comparison between the achievable rates of AMP, and separate optimal MMSE detection and ideal SISO decoding in \cite{Guo2005random, Tanaka2002} with $\beta=N/M=\{1, 1.5\}$, where $C_{\rm Gau}$ denotes the Gaussian capacity and the achievable rates of AMP with Gaussian signaling, $R_{\mr{T-AMP}}$ and $R_{\mr{AMP-DEC}}$ respectively denote the achievable rates of AMP and AMP-DEC with QPSK, 16QAM and 8PSK modulations. }\label{Fig:Rate_AMP_AWGN}
\end{figure} 

Fig. \ref{Fig:Rate_AMP_AWGN} compares AMP and AMP-DEC. For QPSK, 8PSK and 16QAM modulations, the achievable rate of AMP-DEC is lower than that of AMP. This gap increases with $\beta$, but is negligible if $\beta$ is small (e.g. $\beta<0.5$ based on our experimental findings). Furthermore, different from the rate of AMP that always increases with the size of constellation, the rate of AMP-DEC decreases with the increasing of the constellation size for large $\beta$.

\newpage
\section{Simulation Results}
This section provides BER simulations for AMP with optimized irregular LDPC codes. The details of irregular LDPC code optimization for AMP can be found in \cite{Lei2019arXiv}.
  
\subsection{BER Comparison with Irregular and Regular LDPC Codes}\label{Sec:BER_per}  
Fig. \ref{Fig:BER_AMP} provides the BER simulations for an LRMS, in which $\bf{x}$ is generated using optimized irregular LDPC codes \cite{Yuan2008Low, Chung2001} {with code length $=10^5$}. The AMP (see Fig. \ref{Fig:model_coded}) for an optimized LDPC coded LRMS is denoted as ``Opt-Irreg''. The APP decoder is implemented using a standard sum-product decoder. The channel loads are $\beta=\{ 0.5, 1, 2\}$ with $(N, M)=(250, 500), (500, 500)$ and $(500, 250)$, respectively.

To verify the finite-length performance of the irregular LDPC codes with code rate $\approx0.5$, we provide the BER performances of the optimized codes. QPSK modulation is used. The rate of each symbol is $R_{\cal C}\approx1$ bits/symbol, and the sum rate is $R_{sum}\approx N$ bits per channel use. The maximum iteration number is $200\sim700$. Fig.~\ref{Fig:BER_AMP} shows that for all $\beta$, gaps between the BER curves of the codes at $10^{-5}$ and the corresponding Shannon limits are within $0.7 \sim 1$~dB.

To validate the advantage of matching principle, we provide AMP for a standard regular (3, 6) LDPC code (denoted as ``(3, 6)'') \cite{Gallager1962}, and a SISO irregular LDPC code \cite{Chung01} (denoted as ``SISO-Irreg''), corresponding to $R_{\mr{AMP-DEC}}$ discussed in Section \ref{Sec:AMP_DEC}.  As shown in Fig.~\ref{Fig:BER_AMP}, when the BER curves of three systems are at $10^{-5}$, the optimized irregular LDPC codes  have $0.8 \sim 2$~dB performance gains over the un-optimized regular (3, 6) LDPC code for  $\beta=\{0.5, 1, 2\}$, and $0.5 \sim 6$~dB performance gains over ``SISO-Irreg'' for $\beta=\{0.5, 1, 2\}$. These results show that code optimization can provide attractive performance improvement, especially for the large $\beta$. 
\begin{figure}[htb]
  \centering
  \includegraphics[width=8.5cm]{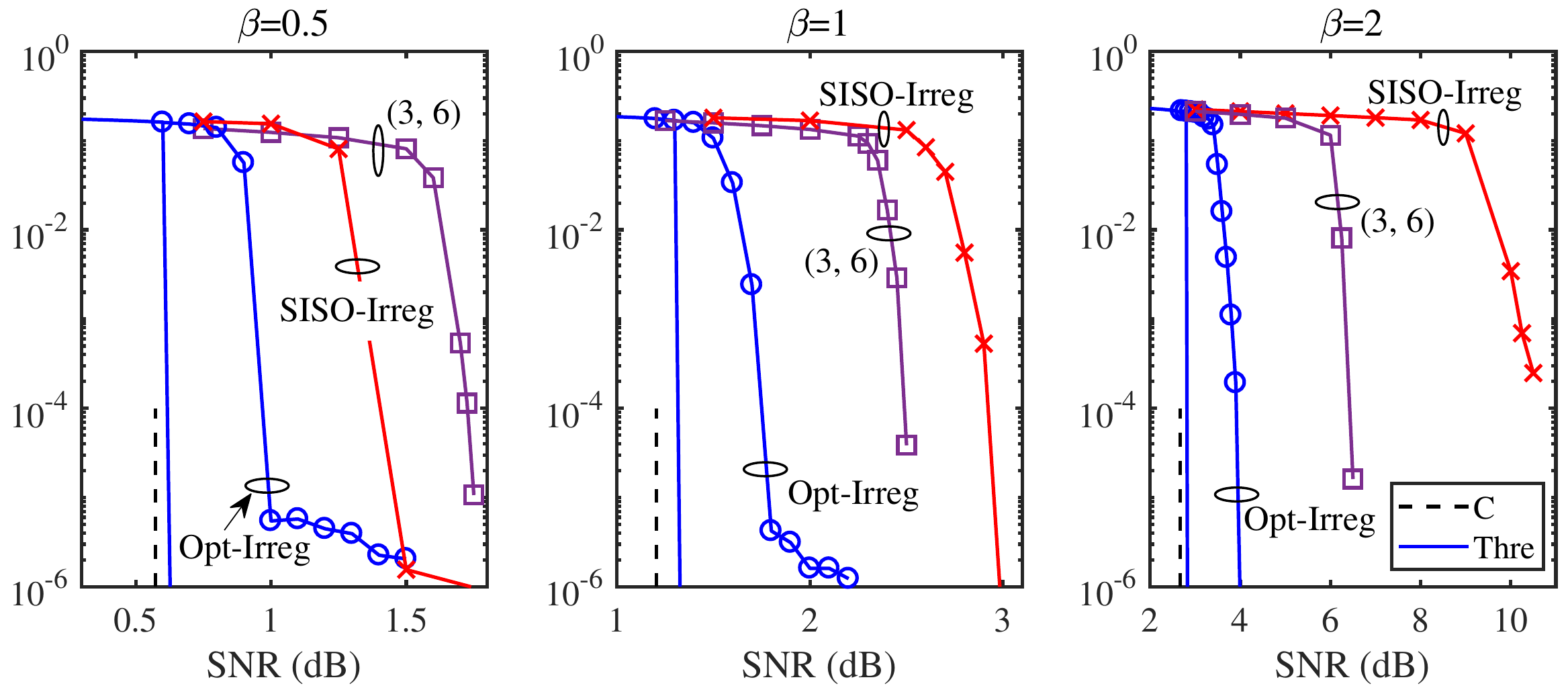}\\\vspace{-2mm}
  \caption{BER performances of AMP, where $C$ deontes the capacity limit, ``Thre''  the BER threshold, ``Opt-Irreg'' the BER of AMP-optimized irregular LDPC codes, ``SISO-Irreg'' the BER with SISO-optimized irregular LDPC codes, ``(3, 6)''  the BER of AMP with regular (3, 6) LDPC code. Code length = $10^5$, code rate $\approx$ 0.5, QPSK modulation, and iterations = $200\sim700$, and $\beta=N/M=\{0.5,1,2\}$. }\label{Fig:BER_AMP}
\end{figure}

\subsection{BER Comparison with Optimized Turbo-LMMSE}
We now compare AMP and Turbo-LMMSE \cite{ YC2018TWC}. We consider a $500\times 333$ QPSK LRMS with $\beta=1.5$. As shown in Fig. \ref{Fig:Rate_AMP_Turbo}(b), the SNR limits of AMP and Turbo-LMMSE for the target rate $R_{\cal C}=1.48\approx1.5$ are $5.38$ dB and $7.99$ dB respectively. Fig. \ref{Fig:BER_AMP_Turbo} shows the BER performances of AMP and Turbo-LMMSE (with iterations $=200$) using optimized LDPC codes. The thresholds of AMP and Turbo-LMMSE are $5.62$ dB and $8.50$ dB respectively, 0.24 dB and 0.51 dB away from the corresponding achievable rate limits, and 0.6 dB and 1.2 dB away from their respective thresholds. We can see that, AMP has 3.5 dB improvement in BER  over Turbo-LMMSE.
\begin{figure}[t]
  \centering
  \includegraphics[width=6.5cm]{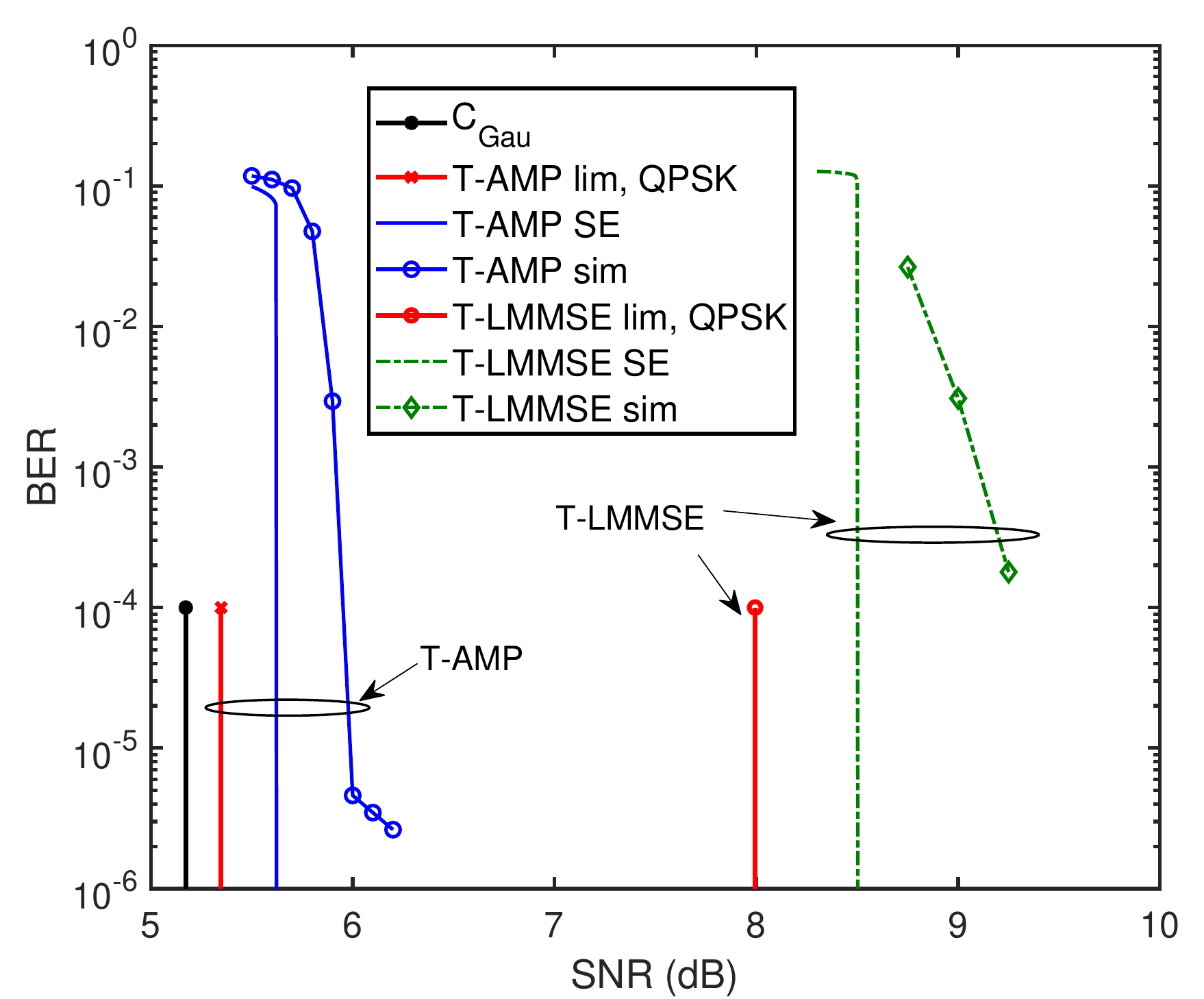}\\
  \caption{BER performances (right) of AMP and Turbo-LMMSE  \cite{YC2018TWC, Lei20161b} with optimized irregular LDPC codes, where $C_{\rm Gau}$  denotes the Gaussian capacity, ``SE''  the state evolution, ``lim''  the QPSK achievable rate limits of AMP/Turbo-LMMSE, ``sim''  the simulated BERs. Code length = $10^5$, code rate $\approx$ 0.74, QPSK modulation,  iterations = $200$,  $N=500$ and $M=333$. }\label{Fig:BER_AMP_Turbo}
\end{figure}

\section{Conclusion}
This paper is on an AMP based scheme for a coded LRMS with arbitrary input distributions. We show that AMP is information theoretically optimal using a curve matching principle and the IIDG assumption. In addition, a code design principle is provided for AMP, and the irregular LDPC codes are considered for binary signaling as an example. The numerical results show that AMP is capacity-approaching (i.e. within 1dB away from the limit) based on optimized irregular LDPC codes, and significant performance improvements ($0.8$ dB $\sim$ $4$ dB) are observed over the system without code optimization. Apart from that, AMP has lower complexity and better performance that the well-known Turbo-LMMSE.


\appendices
\section{Proof of Proposition \ref{The:area_LRMS}}\label{APP:Consistency} 

Let ${\rho}^*  \!=\! snr/(1\!+\!\zeta^*)$, i.e. $\zeta^*\!=\!snr/{\rho}^* \!-\!1$. Then the fixed point function in \eqref{Eqn:dis_cap} is rewritten to 
\BE
snr/{\rho}^* -1 = \beta \, snr \,\omega_{\cal S}({\rho^*} ),
\EE
which is equivalent to the fixed point function $\omega_{\cal S}(\rho)=\phi^{-1}(\rho)$. Substituting  \eqref{Eqn:C_mmse} and ${\rho}^*  = snr/(1+\zeta^*)$ into \eqref{Eqn:dis_cap_new2}, we have\vspace{-2mm}
\BS \begin{align}
&\!\!\!A_{\omega_{\cal C}^*} =  \beta^{-1}\big[\rho^{*}/snr\!-\!\log(\rho^{*}/snr)\!-\!1\big] \!+\!\! \int_0^{\rho^{*}} \!\!\! \omega_{\cal S}(\rho) d \rho\\ 
&\!\!\!= \!\beta^{-1}\!\big[\! \log({1\!+\!\zeta^*})\!-\!{\zeta^*}\!/\!({1\!+\!\zeta^*})\! \big] \!\!+\! C_{\rm SISO} \big(snr\!/\!(1\!+\!\zeta^*)\!\big).
\end{align}\ES 
This is the same as the capacity $C$ in \eqref{Eqn:dis_cap}. Hence, we complete the proof of Proposition \ref{The:area_LRMS}.

\end{document}